# Optimal Dynamic Reconfiguration of Distribution Networks

Rida Fatima
Department of Electrical and Computer Engineering
University of Houston
Houston, TX, USA
rfatima4@uh.edu

Hassan Zahid Butt
Department of Electrical and Computer Engineering
University of Houston
Houston, TX, USA
hbutt@uh.edu

Xingpeng Li
*Senior Member, IEEE*
Department of Electrical and Computer Engineering
University of Houston
Houston, TX, USA
xli82@uh.edu

*Abstract*- **The aim of distribution networks is to meet their local area power demand with maximum reliability. As the electricity consumption tends to increase every year, limited line thermal capacity can lead to network congestion. Continuous development and upgradation of the distribution network is thus required to meet the energy demand, which poses a significant increase in cost. The objective of this research is to analyze distribution network topologies and introduce a topology reconfiguration scheme based on the cost and demand of electricity. Traditional electrical distribution networks are static and inefficient. To make the network active, an optimal dynamic network topology reconfiguration (DNTR) is proposed to control line switching and reconnect some loads to different substations such that the cost of electricity can be minimized. The proposed DNTR strategy was tested on a synthetic radial distribution network with three substations each connecting to an IEEE 13-bus system. Simulation results demonstrated significant cost saving in daily operations of this distribution system.**

*Index Terms*- **Distribution network topology, Dynamic distribution system, Mixed-integer linear programming, Network reconfiguration, Power system flexibility, Topology control.**

## NOMENCLATURE

*Sets:*
| | |
|---|---|
| $I$ | Substations |
| $N$ | Buses |
| $K$ | Non-flexible lines |
| $L$ | Flexible lines |
| $K(n-)$ | Non-Flexible lines of which bus $n$ is the from-bus |
| $K(n+)$ | Non-Flexible lines of which bus $n$ is the to-bus |
| $L(m-)$ | Flexible lines of which bus $m$ is the from-bus |
| $L(m+)$ | Flexible lines of which bus $m$ is the to-bus |
| $I(n)$ | Substation at bus $n$ |
| $T$ | Hours in day |

*Indices:*
| | |
|---|---|
| $i$ | Substation $i$, an element of set $I$ |
| $n$ | Bus $n$, an element of set $N$ |
| $k$ | Non-flexible line k, an element of $K$ |
| $l$ | Flexible line $l$, an element of set $L$ |
| $t$ | Time $t$, an element of set $T$ |

*Parameters:*
| | |
|---|---|
| $C_{it}$ | Cost/price at substation $i$ at time $t$ |
| $D_{nt}$ | Load at bus $n$ at time $t$ |
| $x_k$ | Reactance of non-flexible line $k$ |
| $x_l$ | Reactance of flexible line $l$ |
| $M$ | A very big number |
| $Rating_k$ | Thermal limit of non-flexible line $k$ |
| $Rating_l$ | Thermal limit of flexible lines $l$ |

*Variables:*
| | |
|---|---|
| $\Theta_{nt}$ | Phase angle of bus $n$ at time $t$ |
| $J_l$ | Status variable (binary) of flexible line $l$ |
| $P_{it}$ | Power at substation $i$ at time $t$ |
| $P_{kt}$ | Power flow on non-flexible line $k$ at time $t$ |
| $P_{lt}$ | Power flow on flexible line $l$ at time $t$ |

## I. INTRODUCTION

THE distribution networks (DN) are radially structured with unidirectional power flows to reduce operation and protection complexity and costs [1]. Network reconfiguration involves modifying the topology of a DN in order to enhance the performance of the distribution system. This is achieved by opening normally closed sectionalizing switches and closing normally open tie switches, while adhering to operational constraints of the system. The primary goals of network reconfiguration are to supply power to loads at minimum cost, increase system security and reliability, improve power quality, reduce energy loss, and improve voltage profiles [2]-[4]. In traditional DN reconfiguration, the network is reconfigured once to minimize loss by transitioning from one topology to another. However, as the use of renewable energy resources and demand response programs increase, the behavior of end user profiles becomes more dynamic. This introduces greater variability in load profiles and presents significant challenges for distribution systems [5]-[7]. The main objective of distribution power systems is to deliver electricity to all load points within acceptable levels of quality at the lowest cost possible, which is a conflicting goal since improving the quality and quantity of energy delivered to customers leads to increased investment and operation costs. Therefore, grid planners and operators must strike a balance that considers increasing energy demand and least cost of electricity [8].

DNs are transitioning towards active network topologies with the increasing integration of distributed energy resources (DERs) especially renewable energy such as solar and wind. DERs when connected in the power system can reduce investment and operational cost, improve system reliability and efficiency, by local load balancing and reducing network losses. Another way to minimize the cost of electricity is reconfiguration of distribution network topology (DNT). The DN reconfiguration involves the modification of the DNT via opening and closing of tie switches to sectionalize the network. This is done to minimize cost and power losses within the network [9]. The integration of intermittent renewable sources is facilitated by a flexible power system, especially when their penetration is high. This involves utilizing storage technologies and flexible demand. Additionally, due to the need for suitable location and land availability, storage technologies are usually installed in remote areas. Despite the introduction of large-scale storage technologies, high renewable energy penetration leads to curtailment caused by network congestion, resulting in an increased reliance on local fossil fuel-based generation. Thus, to effectively utilize renewable energy and avoid its wastage, it becomes necessary to implement an efficient smart grid and employ new technologies such as flexible AC transmission Systems (FACTS). To alleviate network congestion and reduce curtailment, transmission expansion planning is done to enhance transfer capabilities [10]. Another approach is to redirect power flow on transmission lines, which can be achieved by adjusting line parameters using FACTS devices [11] or network reconfiguration techniques [12]-[14]. However, expansion planning, energy sources, and FACTS devices demand significant investments and maintenance costs, making reconfiguration an attractive option as it enables the utilization of renewable energy without additional financial commitments. In transmission networks, corrective network reconfiguration is applied to rectify the congestion and transition the system from an emergency state to a normal secure/insecure state. When utilized as a corrective measure, network reconfiguration has the ability to redirect line flows and alleviate congestion in the network following a contingency event. This, in turn, enables low-cost generators to generate additional power. This thus implies that corrective network reconfiguration is exclusively employed when a contingency arises, and the subsequent network becomes overloaded. Moreover, it mitigates the occurrence of significant system disruptions that network reconfiguration may cause, and it also reduces wear and tear on circuit breakers [15]. This network reconfiguration has demonstrated promising outcomes in transmission networks, particularly in corrective transmission switching and the minimization of network congestion [16].

Various network reconfiguration methods have been proposed in literature to address distributed generation integration problems in distribution system, including heuristic algorithms [17]-[19], simulated annealing [20], discrete mutant particle swarm optimization, and binary particle swarm optimization [21]-[22]. Prior research has demonstrated that topology reconfiguration in distribution systems can minimize costs, enhance system reliability, reduce line congestion and support integration of distributed generation. However, most studies have focused on networks based on a single substation and have not considered topology reconfiguration among multiple substations. As distribution networks are transitioning towards more active networks, the potential for regional markets and locational marginal prices (LMP) has increased. Integration of DERs especially renewable energy can significantly impact the LMP [23].

As evident from literature, network topology reconfiguration can play a prominent role in corrective switching in case of contingency events, reduce network congestion, promote integration of DERs in transmission and distribution networks.

The objective of this paper is to explore the potential cost reduction that could be obtained by implementing optimal dynamic network topology reconfiguration (DNTR) among multiple substations based on the LMP at each substation.

The rest of this paper is organized as follows: Section II presents the proposed methodology and mathematical model. Section III describes the assumptions that are considered for the model testing and verification. Section IV analyzes the simulation results of all the test cases, followed by the conclusion and future work in sections V and VI respectively.

## II. MATHEMATICAL MODELING

DNs play a significant role in an electrical power system by creating a linkage between transmission networks and end-users to deliver electricity produced by generation plants. Historically, DNT reconfiguration was considered only in cases of contingencies and network losses reduction. However, with growing energy demand, dynamic energy markets, and integration of renewable energy, it can be utilized to optimize the operational cost of the network as well. This paper implements DNTR to determine the optimal operational cost and topology solution in a day-ahead scheduling under multiple loading levels: a DNTR model with different network topology reconfigurations. DNTR is based on the simplified DC power flow model and is subject to base-case physical requirements of the traditional distribution network constraints while meeting the demand. The mixed integer optimization technique is used to reduce total operational cost in (1),

$$min \sum_{i \in I} \sum_{t \in T} C_{it} * P_{it} \quad (1)$$

where $C_{it}$ is the electricity price at substation $i$ at time $t$; and $P_{it}$ is the imported power at substation $i$ at time $t$. Two sets of lines are considered: (i) flexible lines, which can be connected and disconnected from the network, and (ii) non-flexible lines, which remains in the network all the time unless outages occur. Nodal power balance constraint for a traditional static DN model in each time interval $t$ is enforced in (2).

$$\sum_{i \in I(n)} P_{it} + \sum_{k \in K(n-)} P_{kt} - \sum_{k \in K(n+)} P_{kt} = d_{nt} \ \forall n, t \quad (2)$$

When considering the control over flexible lines in the proposed model, the nodal power balance constraint can be reformulated as shown in (3).

$$\sum_{i \in I(n)} P_{it} + \sum_{k \in K(n-)} P_{kt} - \sum_{k \in K(n+)} P_{kt} + \sum_{l \in L(m-)} P_{lt} - \sum_{l \in L(m+)} P_{lt} = d_{nt} \ \forall n, t \quad (3)$$

Thermal limit constraints for non-flexible and flexible lines are enforced in (4) and (5) respectively.

$$-Rating_k \leq P_{kt} \leq Rating_k \quad (4)$$
$$-J_{lt} Rating_l \leq P_{lt} \leq J_{lt} Rating_l \quad (5)$$

where binary variable $J_{lt}$ is introduced to incorporate line switching in the model and indicate ON/OFF status of flexible lines at time interval $t$. $J_{lt}$=1 indicates that line $l$ is connected to network and $J_{lt}$=0 indicates line $l$ being disconnected from the network. Power flow of the non-flexible lines is calculated by (6), while (7) is used for flexible lines.

$$P_{kt} = \frac{\theta_{kt}}{x_{kt}} \quad (6)$$

$$-BigM(1 - J_{lt}) \leq P_{lt} - \frac{\theta_{lt}}{x_k} \leq BigM(1 - J_{lt}) \quad (7)$$

To ensure the linearity of the system the power flow on the flexible lines is determined by the *Big-M* method [24]. All the lines in a network can only be connected to one substation at any time to avoid loop in a network. To ensure that there is no loop in DN, a constraint is enforced in (8). This constraint will ensure no sub-station is connected to another substation if all the busses have non-zero load. If in reality some buses have zero load, we can assign them small positive numbers such as 0.00001.

$$N_n = N_L + N_s \quad (8)$$

where $N_n$ denotes the number of nodes, $N_L$ is the number of lines in the network, and $N_s$ indicates the number of substations. Equation (9) computes the total lines connected in the system,

$$N_L = N_{FL} + N_{NFL} \quad (9)$$

where $N_{FL}$ is the number of flexible lines and $N_{NFL}$ is the number of non-flexible lines.

## III. TEST CASE DESCRIPTION

In this paper, a three-phased balanced system with a look ahead time of 24 hours is assumed. The test system taken consists of three independent DNs, each constituting a substation connected to the IEEE 13-bus system. The three networks thus contain a total of 3 substations and 39 buses. It is assumed that all three networks have some flexible lines that can connect any two networks together, subject to physical constraints of the network. The switching of flexible lines is affected by the day-ahead LMP at each substation. The LMP at each substation at each time period in a typical day is shown in Fig. 1. It can be seen in the figure that substation-2 has the least cost at most of the hours except between hours 7-8 and hours 20-21. At hours 20 and 21, all three substations have high costs. Overall, substations 1 and 2 have lower LMPs as compared to substation-3.

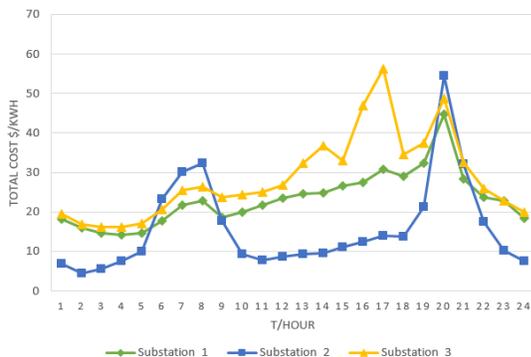
Fig. 1. LMP on each substation in different hours of the day.

Note that the IEEE 13-bus system is modified with respect to line ratings and reactance, and the overall test system is a combination of three modified IEEE 13-bus feeders operating at 4.16kV voltage level [25]. Fig. 2. shows the topology of the base case scenario. In this case, all three substations can be connected through two flexible lines highlighted in red. It is assumed that the flexible lines connecting each substation already exist in the system and thus there is no capital cost associated. It is also assumed that the lines adjacent to the flexible lines are also flexible and thus there are a total of 6 flexible lines in this case.

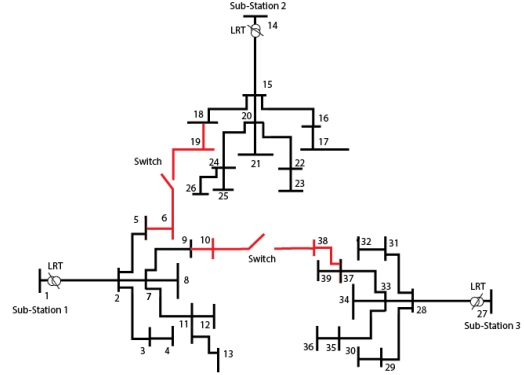
Fig. 2. Base case network topology.

Hourly based actual load profile is considered from Electric Reliability Council of Texas (ERCOT). The load profile represents the pattern of electricity consumption throughout the day, typically in hourly increments. It provides valuable insights into the variations in electricity demand, which can be influenced by factors such as weather conditions, economic activity, and human behavior.

A base normalized load profile is derived by dividing the actual load values at each time interval by the maximum load value observed within the defined time period. Once normalized, the load profile is used to analyze consumption patterns and identify peak demand periods and is shown in Fig. 3. From the pattern given in the figure it can be seen that peak energy demand is between hours 14 to 17.

In the following section, sensitivity analysis of topology reconfiguration is presented.

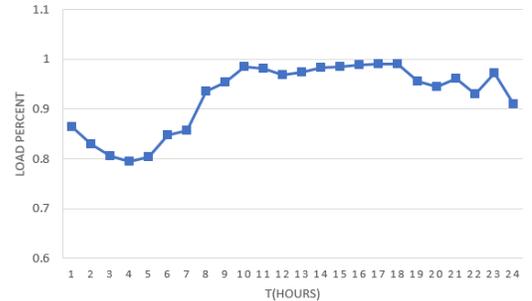
Fig. 3. Hourly normalized percentage load profile.

## IV. CASE STUDIES

### A. DNT-Case I

In case 1, it is assumed that amongst the 39 lines in the system, only 6 lines are flexible that can be connected or disconnected from the system. Rest of the lines are fixed, and reconfiguration of flexible lines will be done based on the LMP in each hour. AMPL software is used to implement the model and Gurobi solver is used to solve the optimization problem. The proposed configuration is tested on three different load scenarios (LS), with LS-1 being the base load case and is referred as 100% load. The load is increased to 110% in LS-2 and 120% in LS-3. The simulation results for DNT-1 are summarized in Table I. It can be seen that the total cost of the system increases as the load increases.

Table I: Simulation results for three different load scenarios

| LS | Total Cost ($) |
|---|---|
| LS-1 | 111,024 |
| LS-2 | 122,126 |
| LS-3 | 133,229 |

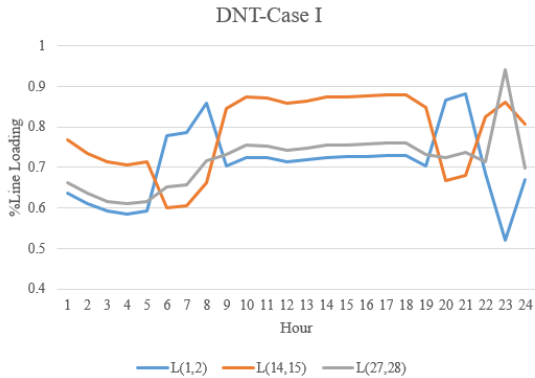

Fig. 4. Percentage line loading of bottleneck lines with LS-1, DNT-1.

Table II: Switching Sequence of Flexible Lines for DNT-1, LS-1

| Hour | L(5,6) | L(6,19) | L(18,19) | L(10,38) | L(9,10) | L(37,38) |
|---|---|---|---|---|---|---|
| 1 | 0 | 1 | 1 | 1 | 1 | 0 |
| 2 | 0 | 1 | 1 | 1 | 1 | 0 |
| 3 | 0 | 1 | 1 | 1 | 1 | 0 |
| 4 | 0 | 1 | 1 | 1 | 1 | 0 |
| 5 | 0 | 1 | 1 | 1 | 1 | 0 |
| 6 | 1 | 1 | 0 | 1 | 1 | 0 |
| 7 | 1 | 1 | 0 | 1 | 1 | 0 |
| 8 | 1 | 1 | 0 | 1 | 1 | 0 |
| 9 | 0 | 1 | 1 | 1 | 1 | 0 |
| 10 | 0 | 1 | 1 | 1 | 1 | 0 |
| 11 | 0 | 1 | 1 | 1 | 1 | 0 |
| 12 | 0 | 1 | 1 | 1 | 1 | 0 |
| 13 | 0 | 1 | 1 | 1 | 1 | 0 |
| 14 | 0 | 1 | 1 | 1 | 1 | 0 |
| 15 | 0 | 1 | 1 | 1 | 1 | 0 |
| 16 | 0 | 1 | 1 | 1 | 1 | 0 |
| 17 | 0 | 1 | 1 | 1 | 1 | 0 |
| 18 | 0 | 1 | 1 | 1 | 1 | 0 |
| 19 | 0 | 1 | 1 | 1 | 1 | 0 |
| 20 | 1 | 1 | 0 | 1 | 1 | 0 |
| 21 | 1 | 1 | 0 | 1 | 1 | 0 |
| 22 | 0 | 1 | 1 | 1 | 1 | 0 |
| 23 | 0 | 1 | 1 | 1 | 0 | 1 |
| 24 | 0 | 1 | 1 | 1 | 1 | 0 |

In DNT- I, sub-station 1 is connected to sub-stations 2 and 3, with six flexible lines. As seen in the LMP graph in Fig. 1, substation 2 is the cheapest one. Thus, the line connecting substation 1 and 2, i.e. line L(6,19) connecting bus 6 and bus 19, is ON at most of the time and line L(5,6) is OFF during the time line L(6,19) is ON so no connection exists between two substations as shown in Table II. By shifting load to the cheapest substation, non-flexible line L(14,15) at substation 2 is heavily loaded most of the time as depicted in Fig. 4. It can also be seen that L(1,2) and L(27,28) are also bottle-neck lines.

At the two sets of time intervals; 7-8 and 21-22, when the LMP at substation 2 is higher than LMP at other substations, the load gets reconnected to substation 1 and thus L(1,2) becomes the heavily loaded line.

### B. DNT-Case II

The network topology for case II is depicted in Fig 5. This case assumes that amongst the 39 lines, 9 are flexible and rest of the lines are not flexible. In this case, all three substations are interconnected with each other unlike previous case. The load from sub-station three can also be shifted to sub-station 2. The proposed configuration is again tested on three different loading factors.

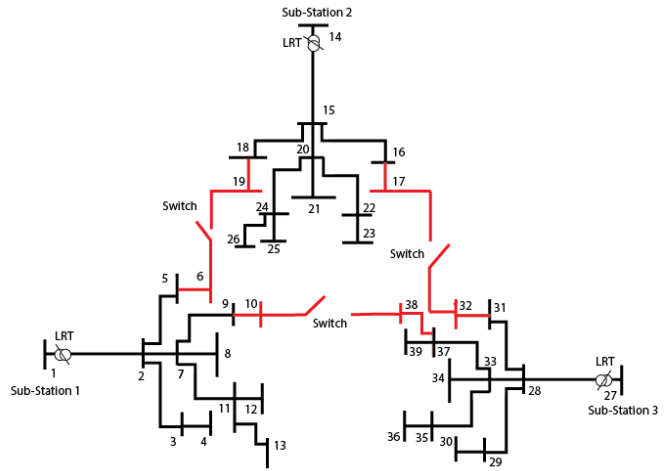

Fig. 5. Network topology with 9 flexible lines.

The simulation results for this case are summarized in Table III, in which it can be seen that the total cost of the system increases as the load increases, similar to the previous case.

Table III: Simulation results for three different load scenarios

| LS | Total Cost ($) |
|---|---|
| LS-1 | 108,054 |
| LS-2 | 118,860 |
| LS-3 | 131,329 |

In this case, because of direct connection of substations 1 and 3 with substation-2 (cheapest substation), L(14,15) is even more heavily loaded as compared to DNT-Case I. Also, it can be seen from Fig. 6 that in the time intervals when LMP is high at substation 2, in addition to L(1,2), L(27,28) also became a bottleneck line.

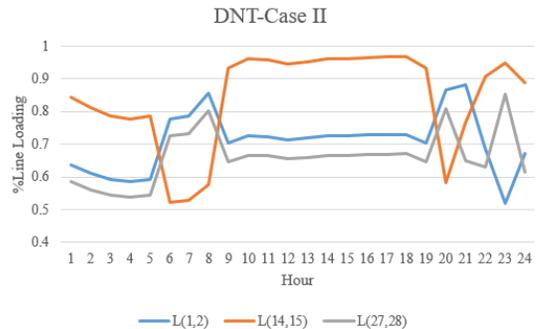

Fig. 6. Percentage line loading of bottleneck lines with LS-1, DNT-II.

## C. DNT-Case III

The network topology for case III is illustrated in Fig. 5. In this case, it is assumed that amongst the 39 lines in a system, 12 lines are flexible and rest of the lines are not flexible. The proposed configuration is again tested on three different loading levels. The simulation results are summarized in Table IV. It can be seen that the total cost of the system increases with the load increases, similar to the previous two cases.

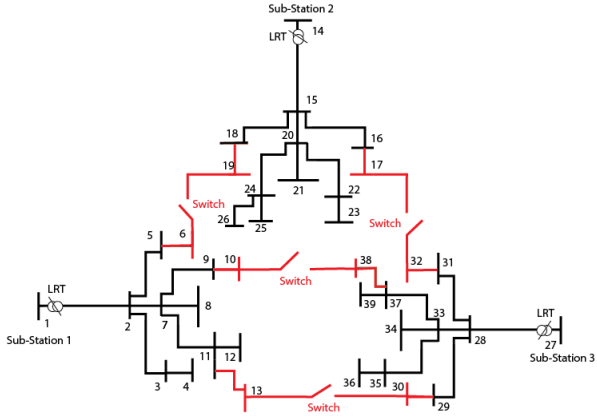

Fig.7. Network topology with 12 flexible lines.

Table IV: Simulation results for three different load scenarios

| LS | Total Cost ($) |
| --- | --- |
| LS-1 | 106,945 |
| LS-2 | 117,653 |
| LS-3 | 130,414 |

In DNT-Case II, all three substations are interconnected with each other like previous case. But in this topology an extra flexible line between node 13 and 30 is introduced to observe the impact of increasing flexible line on operational cost and line loading of the network. From Fig. 8, it can be seen that with increasing number of flexible lines in the network, L(27,28) is not a bottleneck line anymore.

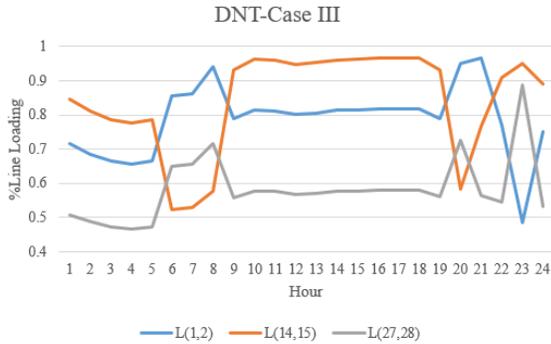

Fig. 8. Percentage line loading of bottleneck lines with LS-1, DNT-III.

## D. DNT-Case IV

In this case, it is assumed that all the lines connected to the network are flexible lines. The topology is same as DNT-Case III except the number of flexible lines have increased from 12 to 40. This proposed topology is again tested on three different loading level. The results show that with increasing loading level, the overall cost of the system increases as evident from the previous three cases as well. Table V summarize the impact of increasing load on operational cost of the system.

Table V: Simulation results for three different load scenarios

| LS | Total Cost ($) |
| --- | --- |
| LS-1 | 100,092 |
| LS-2 | 114,227 |
| LS-3 | 129,615 |

In DNT-IV, as all the lines are flexible, the maximum power is drawn from the cheapest substation. In Fig 9, it can be seen that L(1,2) and L(14,15) are heavily loaded at most of the times and L(27,28) is not a heavily loaded line anymore.

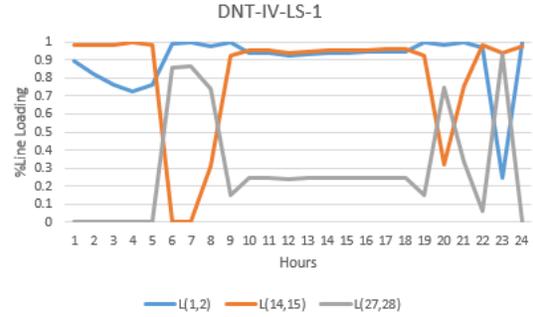

Fig. 9. Percentage line loading of bottleneck lines with LS-1, DNT-IV

Fig. 10 summarizes the results of operational cost reduction from all four cases. It can be observed that by increasing number of flexible lines under the same loading conditions, the total cost of the system is decreasing. For instance, under LS-1, if we increase the number of flexible lines from 6 to 9, there is a total cost reduction of $2,970 per day which is about 2.51% decrease in original operational cost. If we further increase the number of flexible lines from 6 to 12, the total cost will decrease by $4,079 which is around 3.44% of total operational cost and if we increase the number of flexible lines from 6 to 40 the total cost will decrease by $10,932 which is around 10%.

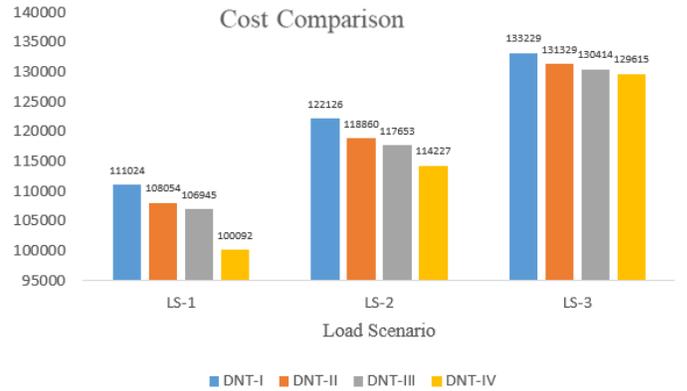

Fig. 10. Comparison of simulation results.

## V. CONCLUSION

In this paper, a network reconfiguration model is presented to optimize the total cost of the system. Presently, the DN is static under normal situations, and it is highly unlikely that a single fixed topology can provide an optimal solution for all scenarios. To address this issue, a day ahead optimal dynamic network reconfiguration strategy is proposed. Mixed integer linear programming is used to formulate the topology reconfiguration problem as a mathematical program with integer decision variables. The results of simulation have shown that by increasing the number of flexible/switchable lines in network topology from 6 to 9, the total cost of the system is reduced by 2.51% as compared to the original operational cost. If we increase the number of flexible lines from 6 to 12, the operational cost would drop by 3.44%. It is also observed that the line loading is significantly impacted by introducing flexible lines in the system.

## VI. Future Work

For DNs, AC based algorithms are needed as DC based algorithms can induce a significant approximation error in the results. Also, large-scale practical systems are needed to demonstrate the robust and effectiveness of the proposed model in contrast to the synthetic 39-bus network tested in this paper. Integration of DERs will be included in the model to observe its effects on operational cost and line congestion.